# Low frequency noise in chemical vapor deposited $MoS_2$

*Yuji Wang [1], Xinhang Luo[1], Ningjiao Zhang[1],Masihhur R. Laskar[1], Lu Ma[2], Yiying Wu[2], Siddharth Rajan[1], and Wu Lu*,[1]*

*[1]Department of Electrical and Computer Engineering,[2]Department of Chemistry and Biochemistry, The Ohio State University, Columbus 43210, U.S.A*

* E-mail: lu.173@osu.edu, Phone: +1 614 292 3462, Fax: +1 614 292 7596

Inherent low frequency noise is a ubiquitous phenomenon, which limits operation and performance of electronic devices and circuits. This limiting factor is very important for nanoscale electronic devices, such as 2D semiconductor devices. In this work, low frequency noise in high mobility single crystal $MoS_2$ grown by chemical vapor deposition (CVD) is investigated. The measured low frequency noise follows an empirical formulation of mobility fluctuations with Hooge's parameter ranging between $1.44\times10^{-3}$ and $3.51\times10^{-2}$. Small variation of Hooge's parameter suggests superior material uniformity and processing control of CVD grown $MoS_2$ devices than reported single-layer $MoS_2$ FET. The extracted Hooge's parameter is one order of magnitude lower than CVD-grown graphene. The Hooge's parameter shows an inverse relationship with the field mobility.

Molybdenum disulphide ($MoS_2$), a layered metal dichalcogenide material, has attracted significant attention recently for next-generation electronics[1], light detection and emission[2,3], and chemical sensing[4] applications due to its unique electrical and optical properties. The intrinsic 2-dimensional nature of carriers in $MoS_2$ offers the advantages of superior vertical scaling for device structure, leading to the potential of low-cost, flexible and transparent 2D electronic devices. Low frequency noise (i.e., 1/f noise), an important characterization parameter of electronic devices, could limit the performance of semiconductor devices. However, there has been little study on monolayer or few layer $MoS_2$ based semiconductor devices. Experiments on graphene multilayers, a 2D material that has been studied extensively, have demonstrated that low frequency noise becomes dominated by the volume noise when the thickness exceeds 7 atomic layers[5]. Below this threshold, the low frequency noise is a surface phenomenon. These results reveal a scaling law for the low frequency noise, which is important for nanoscale devices and for proposed 2D materials applications in sensors, analog circuits, and communications. Since different fluctuation processes can be responsible for low frequency noise in various materials, devices and fabrication processes, a thorough investigation of specific features of low frequency noise in $MoS_2$ is of much interest.

In this work, the low frequency noise of high mobility single crystal $MoS_2$ is investigated by using transmission line measurements (TLM). The measured low frequency noise follows an empirical formulation of mobility fluctuations with Hooge's parameter ranging between $1.44\times10^{-3}$ and $3.51\times10^{-2}$. The Hooge's parameter shows an inverse relationship with the field mobility.

As shown in Fig.1, $MoS_2$ was grown on sapphire by chemical vapor deposition (CVD) with (0001) orientation[6]. The fabrication process started with a thin-film Mo layer deposited on the sapphire base through e-beam evaporation. Then, the sample was in an ambient of sulfur vapor at a high temperature of 900 $^{o}$C for CVD growth. The thin film metal Mo reacted with sulfur vapor and formed $MoS_2$. The thickness or the number of layers of $MoS_2$ can be controlled by accurately controlling the thickness of Mo layer. Subsequently, the fabrication process of TLM pattern began with mesa etching using $Cl_2/O_2$ plasma in an inductively coupled plasma reactive ion etching system. Ti/Au were deposited by e-beam evaporation, patterned by a lift-off process, and annealed at 850 °C for 30 seconds for ohmic contact.

DC characterization was performed using an Agilent 4156C precision semiconductor parameter analyzer. A low noise current lock-in preamplifier SR570 (Stanford Research Systems) and a spectrum analyzer E4440a (Agilent) were used for low frequency noise measurements. The device was biased by a battery to eliminate system noise.

Fig. 2(a) illustrates current voltage (I-V) characteristics of few-layer $MoS_2$ TLM structure in linear scale. The I-V characteristics, shown in a log-log scale (Fig. 2(b)), demonstrate two different regimes: linear relationship between current and voltage at a lower current density, and second-order polynomial relationship at a higher current density. For thin film semiconductor material, the relationship between space charge current density and voltage is

$$J = 2\mu\varepsilon_0\varepsilon_r V^2/(\pi L^2)$$

where $\varepsilon_r$ is the relative permittivity of semiconductor material, $L$ is the length of semiconductor material between contact and $\mu$ is mobility [7]. Once the coefficients of the relationships are determined, the mobility $\mu$ can be estimated independently from each curve. The extracted mobility values are about several tens, up to 72 $cm^2$/Vs.

All fourteen tested devices were selected randomly from one wafer. The low frequency noise measurements manifested two types of behavior according to current noise power spectral density (PSD). The first group of devices (represented by device two shown in Fig. 3(a)) simply follows $1/f^{\beta}$ dependence. The frequency exponent $\beta$ of our devices was found within the range $0.9 < \beta < 1.12$. There are three devices in the second group, which was represented by device one in Fig. 3(a). This type of devices exhibited a broad peak, which could be contributed to generation-recombination (G-R) noise. G-R peaks can be better represented by plotting the product of current noise PSD and frequency versus frequency in a linear-log scale[8]. The peaks are the characteristic frequencies ( $f = (2\pi\tau)^{-1}$ ) of G-R levels, where $\tau$ is the time constant.

One distinct G-R peak was observed in our measurement (as shown in Fig. 3 (b)), at which $\tau$ was $7.96\times10^{-7}$ s .

Fig. 4(a) is the current noise power spectral density (PSD) of a specific spacing with varying voltage applied. As shown in Fig. 4(b), the linear dependence of current noise PSD to $I^2$ (current between two contact pads) suggests that our devices are subject to the mobility fluctuation model in air at room temperature.

The features of low frequency noise are usually illustrated by Hooge's empirical equation [9,10],

$$\frac{S_I(f)}{I^2} = \frac{\propto_H}{f^{\beta} N}$$

where $S_I(f)$ is the current noise PSD, $I$ is device current, $\alpha_H$ is the Hooge's parameter, $f$ is frequency, $N$ is the number of carriers and $\beta$ is the frequency exponent. In this work, our results are interpreted based on the Hooge model. The Hooge's parameter, $\alpha_H$, not necessarily a constant, may instead depend on crystal quality and the scattering mechanisms that determine the mobility $\mu$. The number of carriers, $N$, was extracted from data fitting of linear regime from I-V measurement. Then, the Hooge's parameter, $\alpha_H$, was calculated by using Hooge's empirical equation. $\alpha_H$ varied between $1.44\times10^{-3}$ and $3.51\times10^{-2}$ for all 14 individual few-layer $MoS_2$ devices. The lowest $\alpha_H$ values are comparable to the reported lowest $\alpha_H$ values of single-layer $MoS_2$ transistor[11] ($5.7\times10^{-3}$) fabricated by mechanical exfoliation, suggesting our CVD grown $MoS_2$ has similar high quality to bulk $MoS_2$, also consistent with Raman measurements [6]. Moreover, a much smaller variation of $\alpha_H$ observed compared to that in reported single-layer $MoS_2$ field effect transistor (FET) ($5.7\times10^{-3}$~1.95) [11] suggesting better material uniformity and processing control of our CVD grown $MoS_2$ devices, esp., exfoliation does not allow control over thickness and area of the film[6] and is not suitable for large scale device fabrication. The Hooge's parameters of our few-layer $MoS_2$ devices are about one order of magnitude higher than most exfoliated graphene based devices[12], whereas one order of magnitude lower than CVD grown graphene[12,13]. Fig.5 shows the relationship between Hooge's parameter $\alpha_H$ and mobility $\mu$. The inverse dependence of $\alpha_H$ with $\mu$ is generally to organic thin-film transistors and graphene FETs.

In conclusion, we have investigated the low frequency noise of high mobility single crystal $MoS_2$. The measured low frequency noise follows an empirical formulation of mobility fluctuations with Hooge's parameter ranging between $1.44\times10^{-3}$ and $3.51\times10^{-2}$. A much smaller variation of $\alpha_H$ suggests better material uniformity and processing control of our CVD grown $MoS_2$ devices than reported single-layer $MoS_2$ FET. The extracted Hooge's parameter is one order of magnitude lower than CVD grown graphene. All these information suggests the potential of CVD grown large area $MoS_2$ 2D semiconductor thin films with high material quality for next generation electronic devices. The Hooge's parameter shows an inverse relationship with the field mobility.

This work is partially supported by The Ohio State University Institute for Materials Research.

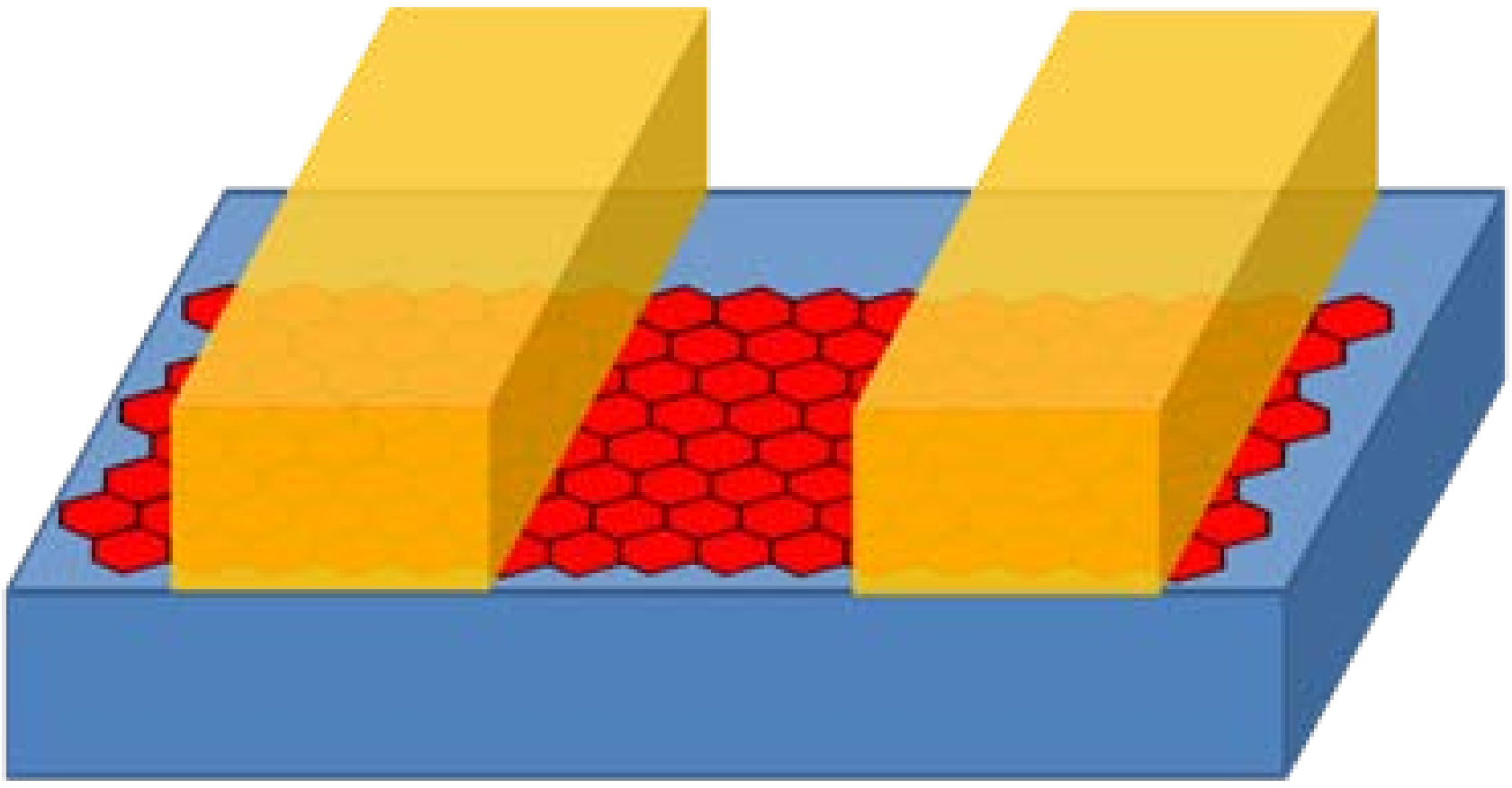

FIG. 1. Device structure. (Blue) substrate: sapphire; (Red) few-layer $MoS_2$; (Gold) Ti/Au contact pads.

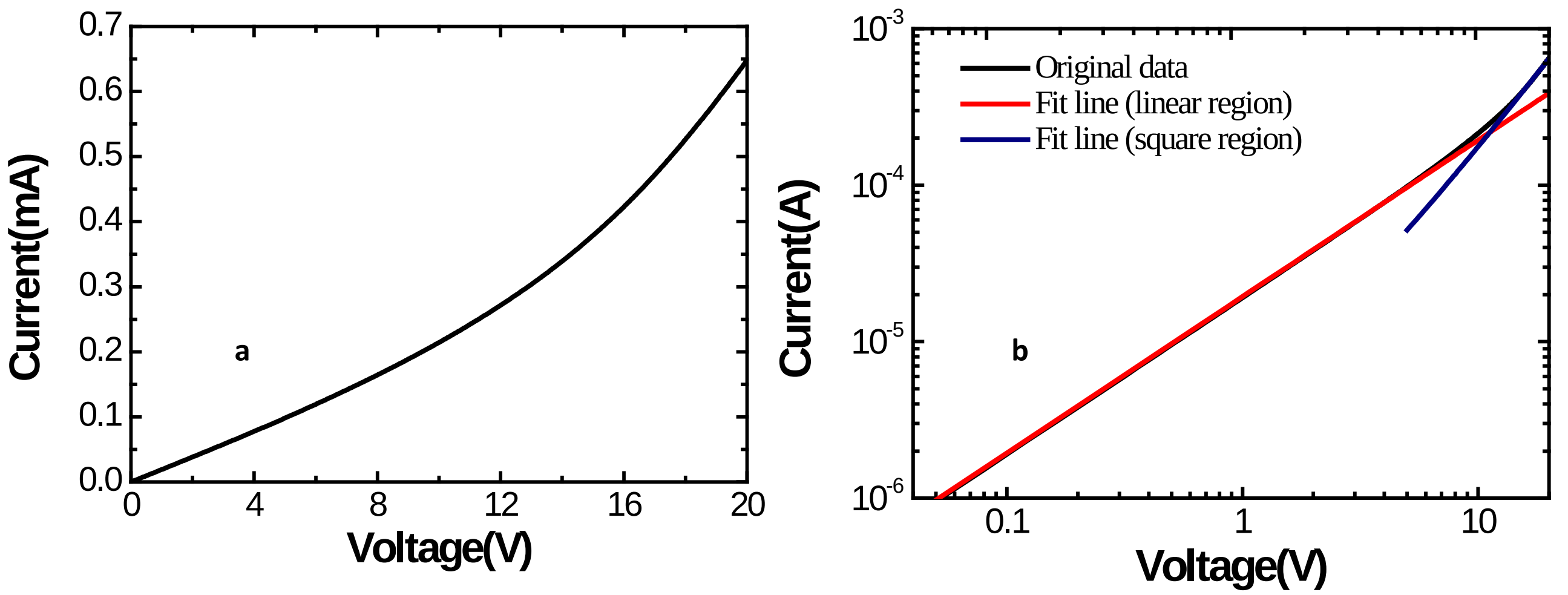

FIG. 2 (a). Current-voltage measurement of 100 μm× 60 μm contact pads with 5 μm spacing. b). The plot in log-log scale for ohmic (linear) and space charge (square) regime fitting.

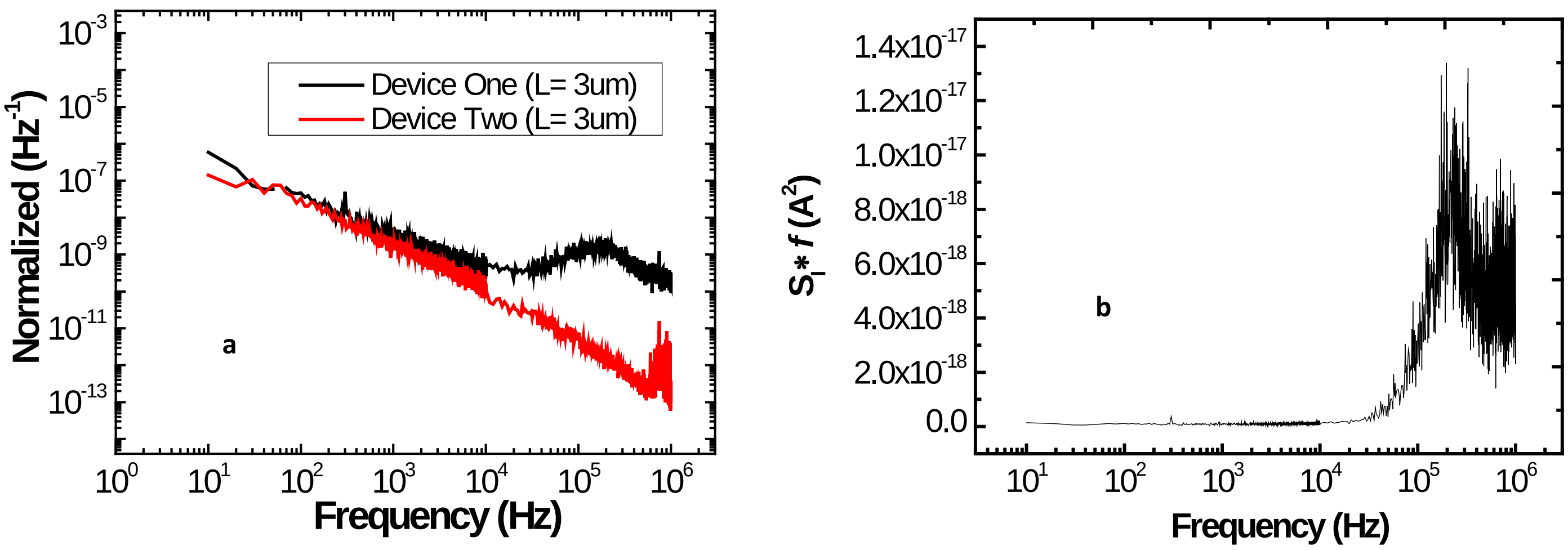

FIG. 3 (a). Current noise power spectral density of two types of devices, represented by two particular devices, labeled by device one and device two. (b). The plot of $S_I \times f$ vs frequency showing a G-R peak for few layer $MoS_2$.

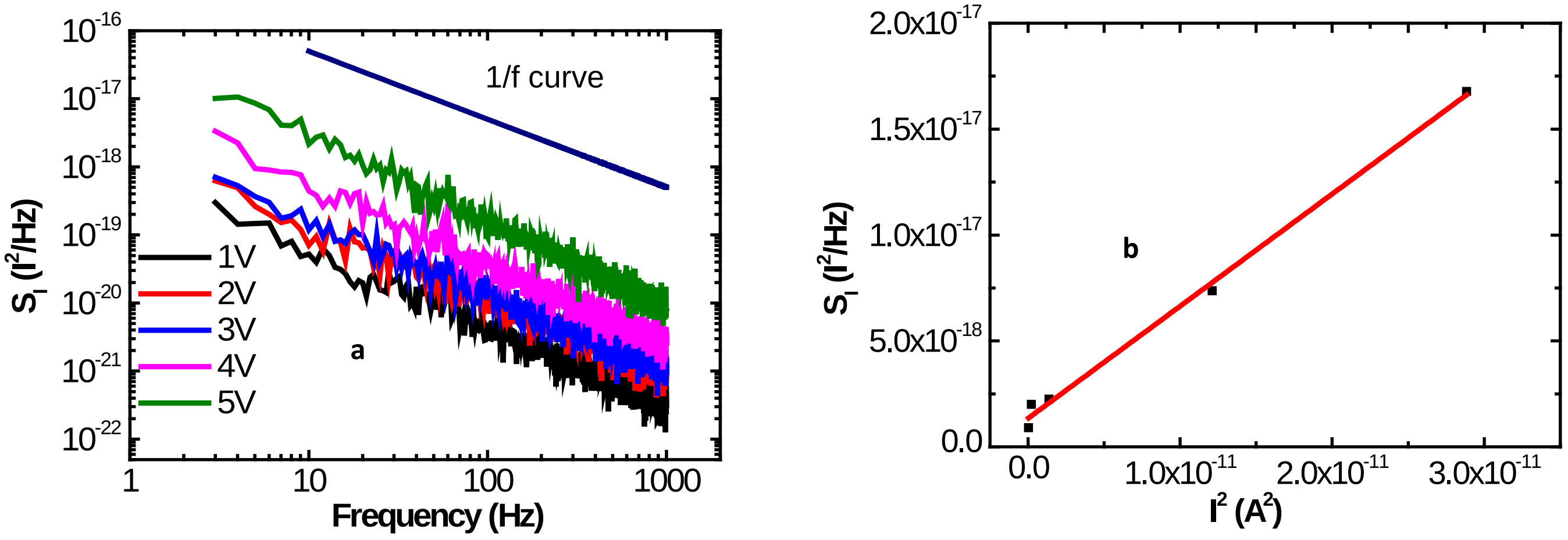

Fig. 4 (a). Current noise power spectral density of a specific spacing with varying voltage applied. (b). Current noise PSD vs. $I^2$.

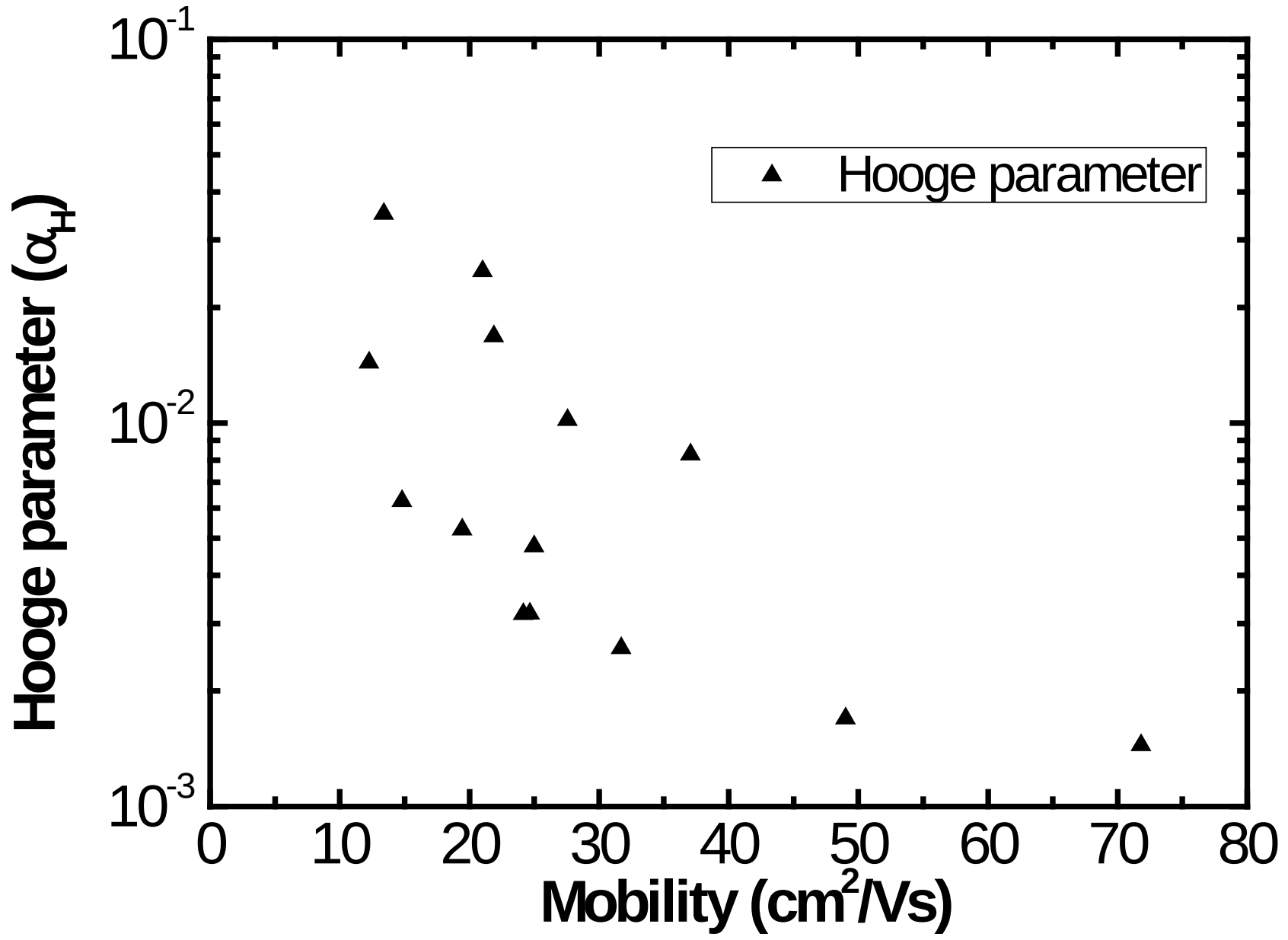

Fig. 5. The Hooge parameter is plotted as a function of mobility for 14 individual devices in air.